# The Past and Future of East Asia to Italy: Nearly Global VLBI

Gabriele Giovannini [1,2,*,†], Yuzhu Cui [3,†], Kazuhiro Hada [4,5,†], Kunwoo Yi [6,†], Hyunwook Ro [7,†], Bong Won Sohn [7,8,9,†], Mieko Takamura [4,10,†], Salvatore Buttaccio [1,†], Filippo D'Ammando [1,†], Marcello Giroletti [1,†], Yoshiaki Hagiwara [11,12,†], Motoki Kino [13,14,†], Evgeniya Kravchenko [15,16,†], Giuseppe Maccaferri [1,†], Alexey Melnikov [17,†], Kotaro Niinuma [18,19,†], Monica Orienti [1,†], Kiyoaki Wajima [8,9,†], Kazunori Akiyama [20,21,22,†], Akihiro Doi [23,24], Do-Young Byun [8,9,†], Tomoya Hirota [4,5,†], Mareki Honma [4,10,†], Taehyun Jung [8,9,†], Hideyuki Kobayashi [14,†], Shoko Koyama [25,26,†], Andrea Melis [27,†], Carlo Migoni [27,†], Yasuhiro Murata [23,24,†], Hiroshi Nagai [5,14,†], Satoko Sawada-Satoh [28,†] and Matteo Stagni [1,†]

1. Istituto Nazionale di Astrofisica Istituto di Radioastronomia, Via P. Gobetti, 101, 40129 Bologna, Italy
2. Dipartimento di Fisica e Astronomia, Bologna University, 40129 Bologna, Italy
3. Research Center for Intelligent Computing Platforms, Zhejiang Laboratory, Hangzhou 311100, China
4. Mizusawa VLBI Observatory, National Astronomical Observatory of Japan, 2-12 Hoshigaoka, Oshu 023-0861, Japan
5. Department of Astronomical Science, The Graduate University for Advanced Studies, SOKENDAI, 2-21-1 Osawa, Mitaka 181-8588, Japan
6. Department of Physics and Astronomy, Seoul National University, Gwanak-ro 1, Gwanak-gu, Seoul 08826, Republic of Korea
7. Department of Astronomy, Yonsei University, Yonsei-ro 50, Seodaemun-gu, Seoul 03722, Republic of Korea
8. Korea Astronomy and Space Science Institute, Daedeok-daero 776, Yuseong-gu, Daejeon 34055, Republic of Korea
9. Department of Astronomy and Space Science, University of Science and Technology, Gajeong-ro 217, Yuseong-gu, Daejeon 34113, Republic of Korea
10. Department of Astronomy, Graduate School of Science, The University of Tokyo, 7-3-1 Hongo, Tokyo 113-0033, Japan
11. Natural Science Laboratory, Toyo University, 5-28-20 Hakusan, Tokyo 112-8606, Japan;
12. ASTRON and Joint Institute VLBI for ERIC (JIVE), Oude Hoogeveensedijk 4, 7991 PD Dwingeloo, The Netherlands
13. Academic Support Center, Kogakuin University of Technology and Engineering, Tokyo 192-0015, Japan
14. National Astronomical Observatory of Japan, 2-21-1 Osawa, Mitaka 181-8588, Japan
15. Moscow Institute of Physics and Technology, Dolgoprudny, Institutsky per., 9, 141700 Moscow, Russia
16. Lebedev Physical Institute, Russian Academy of Sciences, Leninsky Prospekt 53, 119991 Moscow, Russia
17. Institute of Applied Astronomy, Russian Academy of Sciences, Kutuzova Embankment 10, 191187 St. Petersburg, Russia
18. Graduate School of Sciences and Technology for Innovation, Yamaguchi University, Yamaguchi 753-8512, Japan
19. The Research Institute for Time Studies, Yamaguchi University, Yamaguchi 753-8511, Japan
20. National Radio Astronomy Observatory, 520 Edgemont Rd., Charlottesville, VA 22903, USA;
21. Massachusetts Institute of Technology Haystack Observatory, Westford, MA 01886, USA
22. Black Hole Initiative, Harvard University, Cambridge, MA 02138, USA
23. The Institute of Space and Astronautical Science, Japan Aerospace Exploration Agency, 3-1-1 Yoshinodai, Sagamihara 252-5210, Japan
24. Department of Space and Astronautical Science, The Graduate University for Advanced Studies, SOKENDAI, Kanagawa 252-5210, Japan
25. Graduate School of Science and Technology, Niigata University, Niigata 950-2181, Japan
26. Institute of Astronomy & Astrophysics, Academia Sinica, P.O. Box 23-141, Taipei 10617, Taiwan
27. Istituto Nazionale di Astrofisica—Osservatorio Astronomico di Cagliari, Via della Scienza 5, 09047 Selargius, Italy
28. Graduate School of Science, Osaka Metropolitan University, Osaka 599-8531, Japan
* Correspondence: ggiovann@ira.inaf.it
† These authors contributed equally to this work.

**Abstract:** We present here the East Asia to Italy Nearly Global VLBI (EATING VLBI) project. How this project started and the evolution of the international collaboration between Korean, Japanese, and Italian researchers to study compact sources with VLBI observations is reported. Problems related to the synchronization of the very different arrays and technical details of the telescopes involved are presented and discussed. The relatively high observation frequency (22 and 43 GHz) and the long baselines between Italy and East Asia produced high-resolution images. We present example images to demonstrate the typical performance of the EATING VLBI array. The results attracted international researchers and the collaboration is growing, now including Chinese and Russian stations. New in progress projects are discussed and future possibilities with a larger number of telescopes and a better frequency coverage are briefly discussed herein.

## 1. Introduction

The East Asia To Italy: Nearly Global VLBI (EATING VLBI) is a collaboration among the radio astronomy groups of the Italian Istituto Nazionale di Astrofisica (INAF), National Astronomical Observatory of Japan (NAOJ), the Korea Astronomy and Space Institute (KASI) and now all institutions associated with the East Asian VLBI Network (EAVN).

This collaboration started in 2010. The INAF Institute of Radio Astronomy (IRA), the Bologna University, and the Institute of Space and Astronautical Science (ISAS)/Japan Aerospace Exploration Agency (JAXA) submitted a project to the Italian Ministry of Foreign Affairs (MAE) to foster scientific collaboration in radio astronomy between Italy and Japan. There were six principal investigators (PIs), including G. Giovannini and Y. Murata. The project was funded. Thanks to these funds, people from Japan and from Italy started exchanging visits, primarily between Bologna and Tokyo. Visits were very fruitful and the collaboration project started. At that time, the main topic was to start a collaboration in preparation of the launch of the VLBI Space Observatory Programme-2 (VSOP2) satellite and next to study compact active objects with VLBI observations using telescopes in Japan and Italy to increase the sensitivity and angular resolution. After the termination of the VSOP2 project (July 2011), the collaboration project continued with the aim to study active galactic nuclei (AGN) at high resolution and to enhance the collaboration and experience in high-frequency observations.

Other groups joined this project as NAOJ and other people from Universities in Japan, Astrophysics researchers from Korea and East Asia and other INAF observatories/Universities in Italy. The observing array now includes the INAF radio telescopes, the Japanese and Korean joint VLBI network (KVN and VERA array, KaVA), and other telescopes (EAVN).

The aim of this paper was to review the basic elements of this relevant collaboration presenting technical and scientific results to discuss the main obtained results and the possible future developments. An important point of this collaboration is also the friendship among people in Italy and East Asia (mostly Japan, Korea, and China) important from the human point of view but also important to obtain highly scientific results and a fruitful collaboration.

In the next sections, we will give a short history of this project (Section 2), the technical properties of the available array (Section 3), the scientific results (Section 4), and the expected developments in the near future (Section 5).

## 2. Project History

In 2010, thanks to funds of the Italian MAE to improve the scientific collaboration between Italy and Japan, we started an exchange program to give the possibility to researchers from Japan to visit Bologna and to researchers from Italy to visit ISAS/JAXA and NAOJ in Japan.

This project was funded for 5 years and it was a good opportunity to have in Bologna a large group of researchers from Japan visit for one month or even longer. Among them, we quote Hiroshi Nagai, Kazuhiro Hada, Shoko Koyama, Akihiro Doi, and Motoki Kino. Monica Orienti, Gabriele Giovannini, Filippo D'Ammando and Marcello Giroletti also visited ISAS/JAXA and NAOJ for some time.

The main scientific topic discussed was an exchange of the experience of Italian researchers in high-resolution VLBI observations at low/intermediate frequency (L and C bands) including the correlation with gamma-ray emission from AGN, and the experience in space VLBI and high-frequency VLBI observations with VERA from Japanese researchers. In 2012, Kazuhiro Hada obtained a Canon Grant to spend one year in Bologna.

A very important step in these visits was a meeting at the beginning of 2011 with Honma-san at NAOJ where, for the first time, the possibility of common VLBI observations in both Italy and Japan to increase the sensitivity and angular resolution of single arrays was discussed. The possibility of VERA + IRA telescopes VLBI observations started.

In October 2011, the Japanese and Italian governments organized an important meeting in Tokyo to review the strong collaboration present in many fields. Murata san and G. Giovannini had the opportunity to present the astrophysical collaboration and the future project of common observations.

An important result was also the starting of a new project led by H. Nagai, in collaboration with NAOJ and the IRA/INAF-VLBI group, named GENJI (Gamma-ray Emitting Notable-AGNs Monitoring by Japanese VLBI). This program started in 2012 to date with the aim to observe gamma-ray AGNs, using the VERA array at 22 GHz [1]. Such a frequent monitor at high resolution is a unique and very interesting observational project in light of important results from the *Fermi* gamma-ray space observatory.

In October 2012, the first meeting with the name 'EATING VLBI' was held in the CNR research area in Bologna. The meeting provided an opportunity to make this collaboration visible to the VLBI community and to promote joint research using this new VLBI network. On 19 February 2013, the first common VLBI observation was organized between an Italian radio telescope, in Noto, and the VERA array, a few hours of observations of 3C 84 and other calibrator sources to check frequency receivers. The experiment did not yield fringes, as the Noto receiver was not cooled and the sensitivity was too low. Three more testing experiments in April (2) and May failed because of problems in the data format and registration.

In 2014, we investigated a different registration format and correlator mode and we organized a test observation of a few hours on 19 Feb 2015 using VERA in Japan and all three radio telescopes, namely Medicina (Mc), Noto (Nt), and Sardinia (Sr) in Italy. The experiment was carried out at 22 GHz with a 1 Gbps recording rate. Finally, fringes were detected! The result was positive but not as clean as we hoped, but it showed that we were on the right track.

On 25 March 2015, during the EU–Korea Innovation Day at the suggestion of the Italian Embassy in Seoul, information and details of the Italian telescopes were presented and a strong collaboration between Japan, Korea (KaVA), and Italy started. In July 2015, we submitted the proposal for a collaboration between Italy and Korea to the Italian government, which was similar to that obtained for Italy–Japan. This proposal was not funded because of the strong limitation of funds, but it was submitted to a call of the NST for three years and approximately EUR 200 thousand.

On 5 April 2016, a long (10 h) observation using VERA, Mc, Nt, and Sr at 22 GHz produced good fringes! (see Figure 1) and on 23 November 2016 Bong Won Sohn, wrote: *Dear colleagues, I am happy to deliver good news which I just received. Our proposal is selected!*

Thanks to these funds and the starting of EAVN operations, a real EATING activity started adding Italian telescopes to EAVN telescopes.

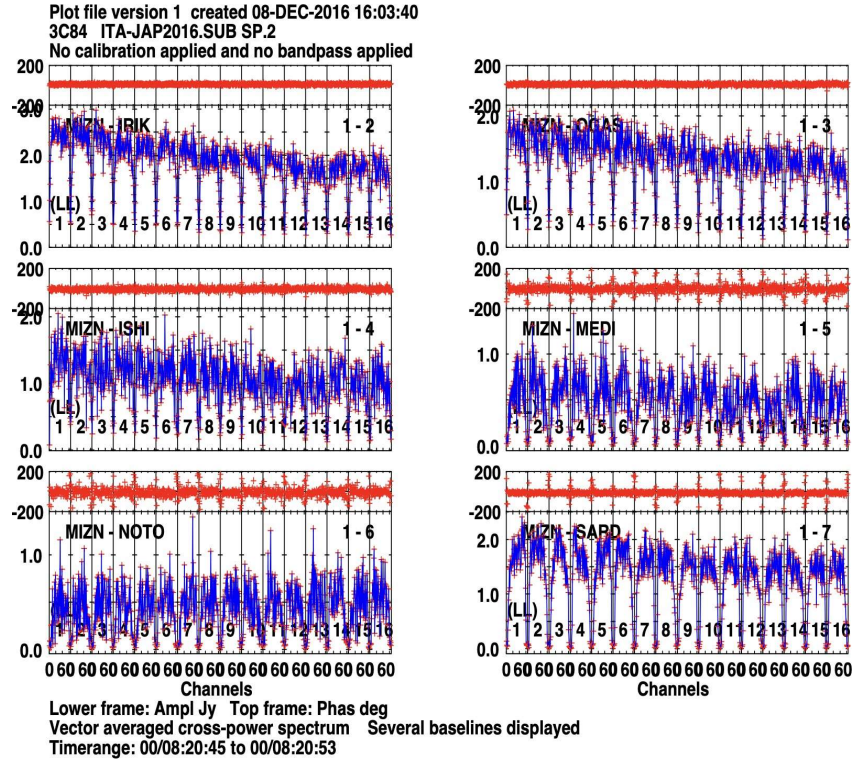

**Figure 1.** Examples of visibility phase and amplitude for a scan of 3C84 in an early VERA-Italy experiment in April 2016.

*EATING Meetings*

- October URL (accessed on 15 January 2023) 2012[1]: Italy, Japan, Korea, Australia—approximately 40 participants, 31 talks;
- October 2014[2]: Italy, Japan, Korea, Australia, Spain, Germany—44 participants, 35 talks;
- October 2017[3]: Italy, Japan, Korea, China, Taiwan—38 participants, 28 talks; URL (accessed on 15 January 2023);
- April 2019[4]: Italy, Japan, Korea, Australia, Spain, Germany, UK, China, Sweden—61 participants, 43 talks. URL (accessed on 15 January 2023).

### 3. Arrays and Correlator

In Table 1, we list the available arrays and their performances, giving reference values for 1 h 1 Gbps observations at 22 GHz. In the following subsections, we provide more information about the single arrays, their combinations, and the correlator facilities.

**Table 1.** List of VLBI arrays and their performances at 22 GHz.

| Array | N. of Stations | 1-h Sensitivity (mJy Beam$^{-1}$) | Angular Resolution Mas × Mas |
|---|---|---|---|
| INAF | 3 | 0.21 | 3 × 2 |
| VERA | 4 | 1.2 | 1.5 × 1 |
| INAF + VERA | 7 | 0.16 | 0.4 × 0.2 |
| INAF + KaVA | 10 | 0.08 | 0.4 × 0.2 |

*3.1. Italian Telescopes*

INAF operates three fully steerable parabolic radio telescopes: two 32 m dishes in Medicina (Mc) and Noto (Nt), and the 64 m Sardinia radio telescope (SRT, or Sr). The locations of the stations are shown in Figure 2 and their coordinates are given in Table 2. The Mc and Nt parabolas were designed and built in the 1980s primarily for VLBI operations in the context of the European VLBI Network (EVN) activities in the cm-wavelength domain. Over the years, they received hardware and software upgrades allowing them to operate with good efficiency, also as stand-alone instruments in continuum [2,3] and spectral lines [4], and up to 43 GHz in Nt thanks to the active surface. The SRT is a more recent project, completed in 2013, and conceived with the goal of operating both as a single dish and as a sensitive VLBI element between 0.3 and 100 GHz [5]. All three stations are connected with broadband optical fibers and can directly send data to the storage units and software correlator in Bologna. The baseline lengths and reference sensitivities are given in Table 3.

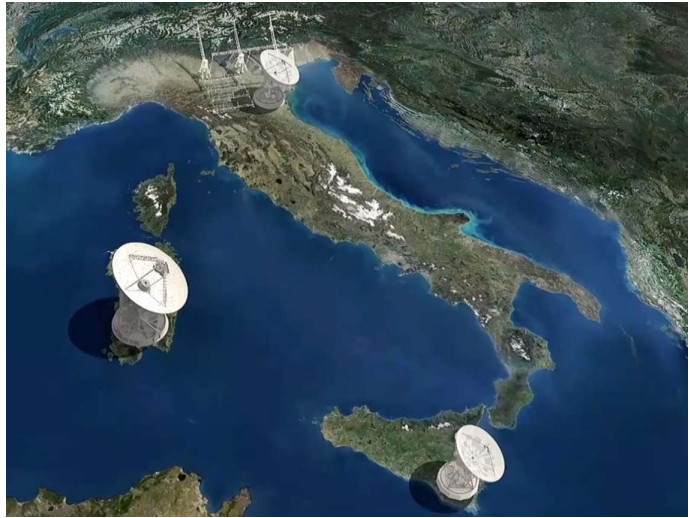

**Figure 2.** The location of the three INAF radio telescopes. From North to South: Medicina, Sardinia, and Noto.

**Table 2.** INAF radio telescopes.

| Location | Code | Size | Year | Latitude | Longitude | Elevation | Operating Receivers |
|---|---|---|---|---|---|---|---|
| Medicina | Mc | 32 m | 1983 | 44° 31′ 15″ N | 11° 38′ 49″ E | 25 m | L, S, C, C+, X, K |
| Noto | Nt | 32 m | 1988 | 36° 52′ 34″ N | 14° 59′ 21″ E | 78 m | P, L, C+, K, Q |
| Sardinia | Sr | 64 m | 2011 | 39° 29′ 34″ N | 9° 14′ 42″ E | 600 m | L, S, C, C+, X, K |

**Table 3.** INAF VLBI baseline length and sensitivity.

| | | Sensitivity (mJy min$^{-1}$ Gbps$^{-1}$) | | |
|---|---|---|---|---|
| Baseline | Length (km) | L-Band | C+-Band | K-Band |
| Mc-Nt | 893 | 5.9 | 7.8 | 6.1 |
| Mc-Sr | 592 | 1.8 | 1.7 | 2.5 |
| Nt-Sr | 580 | 1.8 | 1.9 | 2.7 |

The three stations are committed to observe the context of the EVN consortium activities for up to □70 days per year. The remaining time is devoted to maintenance and development, single-dish activities, geodetic observations in the framework of the International VLBI Service (IVS, for Mc and Nt), support for the Italian Space Agency (ASI, for Sr), and other VLBI projects that can be arranged on an ad hoc basis, such as the recent PRECISE[5] project (Pinpointing REpeating ChIme Sources with EVN dishes) devoted to the localization of fast radio bursts (FRB, [6]). Observation times are offered twice per year through open calls for proposals[6], and the allocation process is managed by an international panel. This has offered the possibility to carry out pilot experiments in coordination with

East Asian partners, and ultimately resulted in a Memorandum of Understanding (MoU) between INAF and KASI, reserving up to 30 h per semester for projects involving the use of INAF and KVN stations, with PIs affiliated with either one of the two institutions (see Section 3.3).

The current suite of receivers for each station is indicated in Table 2. However, the situation is in rapid evolution. Driven by several ambitious scientific goals (including compliance with the latest EVN roadmap [7] and the possibility to further develop joint observations with East Asian partners), INAF has obtained substantial financial support through the National Operative Program of the Italian Ministry of University and Research. With this program, INAF recently purchased compact triple-band receivers, developed and built in Korea, to perform simultaneous VLBI observations in the K-, Q-, and W- bands (22/43/86 GHz) with all three of its radio telescopes. Within the same project, funding is available to improve the surface accuracy in Medicina (to allow W-band observations) and the Italian VLBI software correlator. Work is in progress, and first test observations could start in spring 2023.

*3.2. KVN and VERA Array (KaVA) and East Asian VLBI Network (EAVN)*

KaVA is a VLBI array that combines KVN and VERA. Its baseline lengths range from 305 to 2270 km [8]. The observation frequencies are 22 and 43 GHz with the maximum angular resolution of 1.24 mas and 0.63 mas at each frequency. The combined KaVA compensates for the limitations of each array, namely the low spatial resolution of KVN and the low sensitivity to extended structure of VERA [8,9]. The operation format of KaVA was naturally inherited to EAVN when the latter became operational [10].

EAVN is an international VLBI facility in East Asia and it is operated under mutual collaboration between East Asian countries, as well as part of Southeast Asian and European countries. EAVN consists of 16 telescopes (4 in China, 8 in Japan, and 4 in Korea) and 3 correlator sites (Shanghai in China, Mizusawa in Japan, and Daejeon in Korea) in East Asia [10,11]. EAVN is currently operated in three frequency bands, 6.7, 22, and 43 GHz with the maximum baseline length of 5078 km between the Urumqi and Ogasawara telescopes, corresponding to an angular resolution of 1.82, 0.55, and 0.28 mas at 6.7, 22, and 43 GHz, respectively. For more details of EAVN, see [11].

EAVN provides part of its observing time as 'the EATING VLBI' session from the second half of 2020 with the maximum of total observation time of 30 h in a semester (see next Section). As of the first half of 2022, thirty-seven-epoch EATING VLBI observations were conducted with the total observing time of 275 h, whereas Italian telescopes joined those sessions with 135 h. A major limitation is the limited source common visibility from Italian and East Asia telescopes.

*3.3. EATING VLBI*

East Asia To Italy: Nearly Global VLBI (EATING VLBI) is a Eurasian VLBI collaboration of East Asian VLBI Network and the Italian INAF radio telescopes [12] (Figure 3). EATING VLBI launched its operation at 22 GHz with a bandwidth of 256 MHz (16 MHz × 16 IF or 32 MHz × 8 IF) and will be operational at 43 GHz soon. KaVA + INAF image sensitivity is 0.08 mJy/beam at 22 GHz (1 Gbps and 1 h integration). The further extension of frequency, polarization, and simultaneous multi-frequency observation capabilities will be available in the next few years. In Figure 4, we show simulated the uv-coverage of EATING VLBI array towards sources at various declination angles. In 2019, an MoU was signed between INAF and KASI for coordinated proposal calls. On 9 April 2019, a Memorandum of Agreement was signed by INAF and KASI which recognized the mutual interest in conducting VLBI experiments, and long-standing collaboration and collaborative research on supermassive black holes. Both parties agreed upon providing 30 h of observation time per semester for common VLBI observations with the INAF telescopes and the KVN. Data are correlated at the Daejeon correlator with the partial support of the DIFX software correlator available in Bologna at INAF. INAF telescope data are transferred by optical fiber to KASI. However, although this MoU does not guarantee the full participation of EAVN, the proposals so far have been positively reviewed by the EAVN Time Allocation Committee when the proposals are submitted to EAVN.

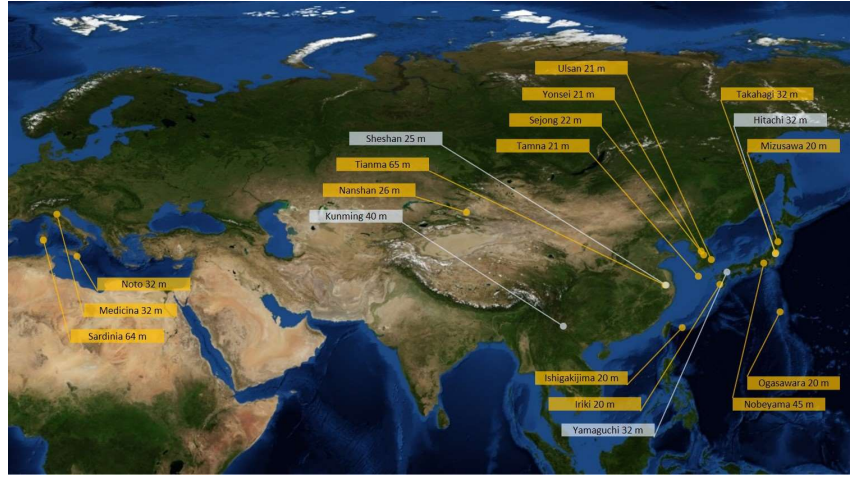

**Figure 3.** Locations of telescopes consisting of EATING VLBI including 19 telescopes (16 EAVN telescopes and 3 Italian telescopes). Telescope names and locations are overlaid on 'the Blue Marble' image (credit of the ground image: NASA Earth Observatory (https://earthobservatory.nasa.gov/, 15 January 2023).

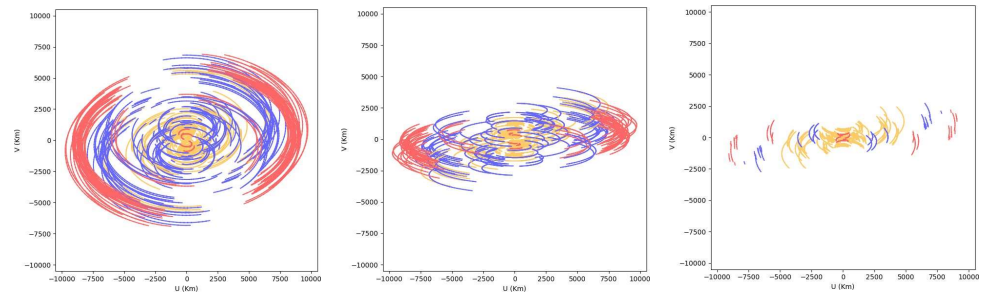

**Figure 4.** Simulated UV-coverages of EATING VLBI towards sources at declinations of $+40°$, $+12°$ and $-30°$, respectively. Curves with orange and red indicate baselines related to EAVN and Italian stations, respectively. Additionally, here we also include 3 Russian stations (Badary, Svetloe, Zelenchuk, with blue-colored curves) that occasionally join EATING VLBI sessions from 2020.

*3.4. Correlators*

3.4.1. The Bologna Correlator

The Bologna correlator is currently made of 3 storage and 2 computing nodes. The mixed storage and computing nodes have a scratch 11 TB SSD storage and 100 TB SAS disk storage, whereas the computing nodes have a 14 TB scratch storage. There is an additional dedicated storage node with 200 TB available. With the current setup, we are able to correlate data sampled at 1 Gbit/s at roughly half of the observation time. The nodes are interconnected via infiniband 40 Gbit and have a 10 Gbit connection to the National research and education network (NREN) coordinated by 'Gruppo per l'Armonizzazione della Rete della Ricerca' (GARR). The planned future upgrades will be to acquire further computing nodes with NVME and SSD storage and integrate the current disk storage into a network file system such as lustre.

### 3.4.2. The Daejeon Correlator at Korea–Japan Correlation Center

The Daejeon Correlator of the Korea–Japan Correlation Center (KJCC) at KASI is a hardware FX correlator which is jointly funded by KASI and NAOJ. It currently processes KVN's international collaboration observations, including the East Asian VLBI Network (EAVN), its precursor KaVA, and the EATING VLBI Network. A maximum of 16 stations (single polarization mode) and a maximum of 8192 Mbps input data rates per station can be processed. Several VLBI data playbacks can be used, such as Mark5B, VERA2000, and OCTADISK. In order to process data from these various playback systems, the Daejeon Correlator uses an intermediate buffer system called the Raw VLBI Data Buffer (RVDB) which is developed by NAOJ. At 22 GHz, KJCC has 0.05 km s$^{-1}$ velocity resolution, ±36,000 km maximum delay and 1.075 kHz maximum fringe tracking[7].

All data have been transferred by optical fiber connection.

## 4. Key Targets and Examples of EATING Images

Here, we present example EATING VLBI images on various sources to demonstrate the typical performance of the EATING VLBI array. Here, we select the data from two representative sessions, for which the observations were largely successful over the array. These are part of ongoing EATING VLBI monitoring programs on M 87 (Session A) and 1H 0323+342 (Session B), for which dedicated studies and detailed data analysis procedures will be published in separate papers. The basic information of these two sessions is summarized in Table 4. In what follows, we describe brief notes and prospects on some key targets of the EATING VLBI program. To produce these images, initial calibration was performed in AIPS, and subsequent self-calibration and imaging in DIFMAP.

**Table 4.** EATING VLBI observations presented in this article.

| Session | Date | Station | Observed Sources |
| --- | --- | --- | --- |
| Session A | 6 Dec 2019 | KVN, VERA, Nanshan, Sardinia | M 87 (primary), 3C273, 3C279, 1219+044 |
| Session B | 1 Dec 20194 | KVN, VERA, Tianma, Medicina, Sardinia | 1H 0323+342 (primary), 3C84, 0250+320, 3C111, 3C120, B0218+357 |

### 4.1. M 87

M 87, a giant elliptical galaxy at a distance of 16.7 Mpc [13], shows a prominent jet up to several kpc away from the center of the galaxy, which can be observed at radio, optical, and X-rays. Since its discovery [14], the jet of M 87 has been extensively monitored over a century. Recently, the Event Horizon Telescope (EHT) Collaboration has successfully imaged the black hole shadow of M 87 and constrained its mass to be $M_{\rm BH} \simeq 6.5 \times 10^9 M_0$ [15]. Beyond the horizon scales, the Global Millimeter-VLBI Array (GMVA) observations at 86 GHz revealed that the jet is limb-brightened, suggesting multiple layers in the jet [16,17]. On scales beyond $\sim$100 Schwarzschild radii ($R_s = 2GM/c^2$), the M 87 jet has been intensively studied with centimeter-VLBI facilities such as VLBA, EVN, and KaVA/EAVN. There is growing evidence that the jet is gradually accelerated over de-projected distances between $\sim 10^2 R_s$ and $\sim 10^6 R_s$ from the jet base [18–22]. The acceleration region is characterized by a parabolic shape [23,24], indicating that jet acceleration and collimation are intimately related as predicted by the magneto-hydrodynamic (MHD) acceleration models. Nevertheless, the velocity profile of the M 87 jet appears to be more complicated. For example, the extrapolation of observed velocities at distances <$10^2 R_s$ expects subluminal motions, while GRMHD simulations [24,25] predict superluminal speeds. Furthermore, the jet acceleration is less efficient compared to the prediction from a highly magnetized jet, which is suggested by other previous observations [17,26]. Indeed, there is still a substantial discrepancy between the observations and our understanding.

In order to overcome the previous limitations, it is required to expand the M 87 monitoring with a high angular resolution, sensitivity, and an observing cadence. EATING VLBI at 22 GHz is currently the only facility that allows us to regularly monitor the innermost jet regions where the initial acceleration takes place. In fact, the angular resolution of EATING along the M 87 jet direction is 0.15 mas ($\sim 67 R_s$) on the sky, which is suitable for reliably investigating the region within $<10^2 R_s$. The first EATING VLBI experiment towards M 87 was performed in 2017 [27]. From the end of 2019, we started our regular EATING monitoring program on a long-term basis. Our preliminary analysis of inner 1 mas kinematics from multi-epoch images detected possible superluminal jet features, but it is not conclusive due to the limited number of epochs and cadence. More detailed analysis is on-going with a typical observation cadence of $\sim$3 week, and the future monitoring will further allow us to better understand the M 87 jet physics.

*4.2. 1H 0323+342*

1H 0323+342 is a well-known Narrow-line Seyfert 1 galaxy (NLSy1). This source is known as the nearest (z = 0.063) γ-ray detected NLSy1s [28] and is one of the very few NLSy1s where the host galaxy is resolved [29]. Thus, this source is a unique target that allows us to probe the vicinity of the central engine at the highest linear resolution among γ-ray active NLSy1s (1 mas = 1.2 pc). The pc-scale structure of this source was first examined by [30]. More recently, an extensive VLBA analysis by [31] discovered a parabolic shape of the inner jet, resolving a collimation zone near the jet base. In addition, the collimation zone ends up with a bright quasi-stationary component "S" at 7 mas from the core, indicating that S represents a recollimation shock that could be associated with the site of high energy γ-ray emission [31].

Nevertheless, the structure and dynamics of the innermost jet regions is still controversial. While the previous multi-epoch VLBA monitoring of this jet reported a highly superluminal motion up to $\sim 9\,c$ [32,33], some of the previous studies claim the possible presence of another stationary feature at 0.3–0.7 mas from the core [30,34]. If this is real, this upstream feature might represent the first recollimation shock rather than S and propose a new site of γ-ray emission. However, previous VLBI images were not conclusive due to insufficient resolution (at ≤22 GHz) or sensitivity (at 43 GHz).

To better probe the innermost regions, EATING VLBI at 22 GHz is an optimal approach thanks to its high resolution, high sensitivity, and high-cadense monitoring capability. From October 2019, we started performing the EATING monitoring observations of this source at typical intervals of 3 months, completing 11 sessions to date. Through this program, the innermost regions of 1H 0323+342 were robustly resolved at 0.15 mas ($\sim$0.18 pc) scales (Figure 5). Remarkably, the EATING images unambiguously detected a bright feature located at $\sim$0.5–0.9 mas from the core, and our preliminary analysis of the multi-epoch images indicate the quite stationary nature of this feature. Therefore, we are indeed beginning to constrain the innermost structure of this source. More detailed analysis is underway and ongoing monitoring will further allow us to test the MHD jet acceleration/collimation paradigm and the Doppler boosting and Lorentz factors that are key parameters to constrain the site of γ-ray emission.

**Table 5.** Summary of image parameters of Figures 5 and 6.

| Session | Source | Beam Size (mas, mas, deg) | $I_{rms}$ (mJy/beam) | $I_{peak}$ (mJy/beam) |
|---|---|---|---|---|
| Session A | 3C279 | 1.61, 0.23, 3.39 | 13.5 | 10267 |
| | M87 | 1.48, 0.33, 9.92 | 0.10 | 769 |
| | 1219+044 | 1.60, 0.27, 6.41 | 0.12 | 1078 |
| | 3C273 | 1.58, 0.23, 5.05 | 7.5 | 5302 |
| Session B | 1H 0323+342 | 0.92, 0.25, 10.1 | 0.12 | 332 |
| | 3C84 | 0.86, 0.25, 15.3 | 12 | 6840 |
| | 0218+35 | 0.85, 0.23, 6.54 | 0.57 | 133 |
| | 3C120 | 1.83, 0.22, 4.54 | 1.96 | 1410 |
| | 0250+320 | 0.99, 0.23, 7.23 | 0.47 | 321 |
| | 3C111 | 0.93, 0.26, 13.8 | 0.80 | 536 |

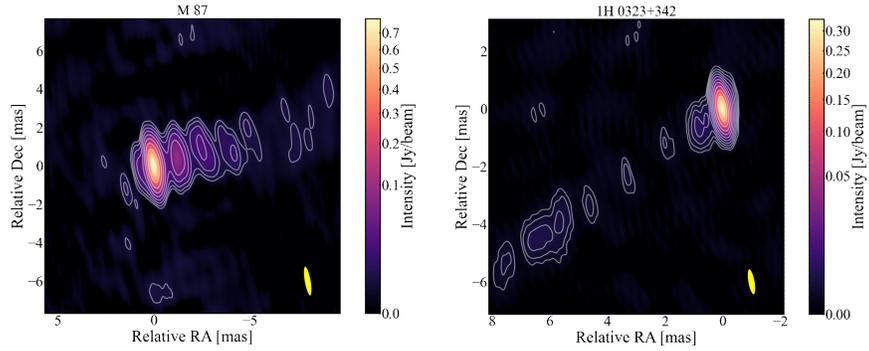

**Figure 5.** Natural weighted images obtained from EATING VLBI: (**left**) M 87 image. (**right**) 1H 0323+342 image. For each image, contours start from 1, 2, 4 ... times a 3$\sigma$ image rms level and increase by factors of 2, and the color bar shows the intensity. The yellow ellipse in the bottom is the beam size of the each map. Image parameters are reported in Table 5.

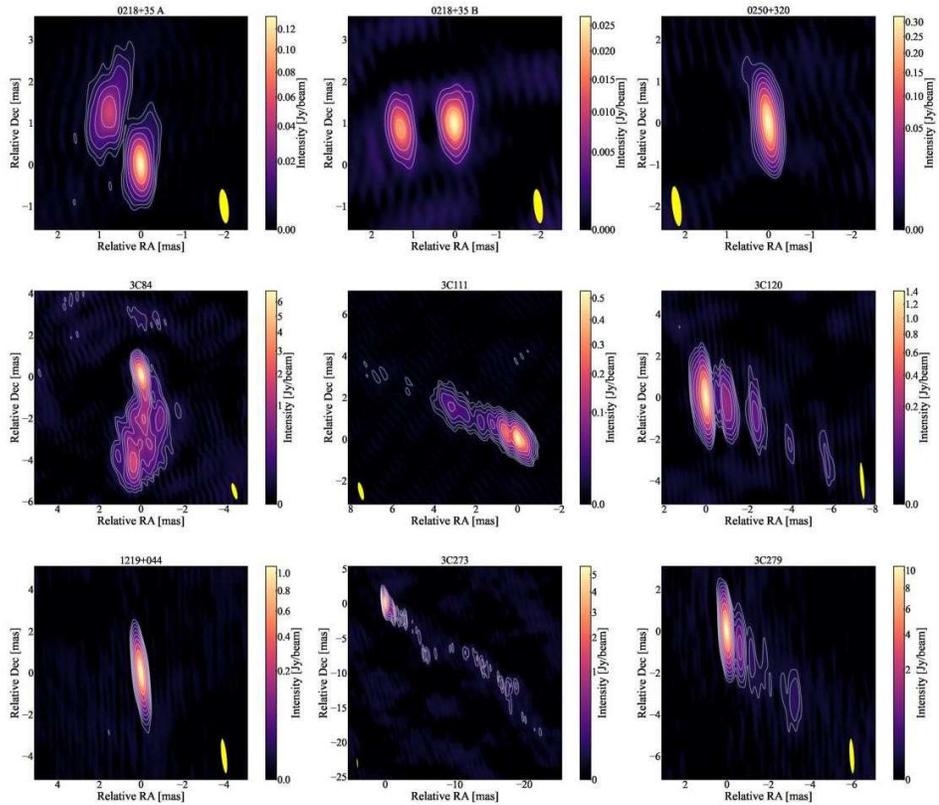

**Figure 6.** This figure shows a summary of the natural weighted images obtained from EATING VLBI. For each image, contours start from 1, 2, 4 ... times a 3$\sigma$ image rms level and increase by factors of 2, and the color bar shows the intensity. The yellow ellipse in the bottom is the beam size of the each map. Image parameters are reported in Table 5.

*4.3. 3C84*

The nearby radio galaxy 3C 84 ($z = 0.0176$) at the center of the Perseus cluster is known as an excellent laboratory for exploring the physics of energy transport by radio lobes at parsec scales. At approximately 2005, it was reported that the radio flux started to increase again [35]. VLBI observations revealed that this flare originated from within the central pc-scale core, accompanying the ejection of a new jet component known as C3 expanding in the Southern direction with respect to the core [36,37]. This new component appeared from the south of the core at approximately 2003 and it propagates southward and becomes brighter [38].

EATING VLBI, with its higher spatial resolution in the east–west (EW) direction, has the advantage of reproducing the fine structure of this object. While physical properties around the C3 component are discussed based on the simple kinematics of the brightness peak of C3 when using VERA and KaVA [36,38–40], the new EATING image shows a spatially resolved fine structure around the brightness peak (Figure 6). This is a good example of the advantage of EATING over VERA and KaVA. The high spatial resolution also allows us to resolve the core component. The EATING image shows that the core is elongated towards the EW direction, more than the beam-size, which is consistent with the double nuclear structure in core region [41]. EATING also has the advantage in imaging extended structures thanks to EAVN-originated short baselines. In contemporaneous data, VLBA did not adequately capture the structure of the entire extended radio lobe [40]. In contrast, EATING was able to capture the overall structure of this newborn radio lobe and its internal sub-structures.

A counter radio lobe on the north side is also shown in the image. This structure is spatially resolved in agreement with its internal structure in the east–west direction [42,43]. Interestingly, the east–west structure of the counter-jet appears to be different when compared to archival images from the Boston Univ. blazar monitoring of the same period: November 2019 (see https://www.bu.edu/blazars/research.html, 15 January 2023). This may suggest changes in the apparent structure of the counter-jet on a short-time scale. This would be an interesting future research topic.

*4.4. B0218+357*

B0218+357 is a distant (z = 0.944) flat spectrum radio quasar (FSRQ) gravitationally lensed by a foreground (z = 0.685) spiral galaxy B0218+357G. The lensing effect splits the AGN into two lensed images (A and B) separated by 335 mas [44]. The source is also known as one of the few gravitationally lensed quasars where active gamma-ray emission is detected up to GeV/TeV energy bands [45,46], rendering this source a unique target with which one can probe the innermost active regions of AGN via microlensing.

From 2017, we started detailed monitoring observations on B0218+357 with KaVA to better constrain the innermost jet structure and its evolution. The initial results are published in [47], where we robustly detected a core–jet structure up to 22/43/86 GHz (KVN only at 86 GHz). Our KaVA monitoring observations were also conducted as part of large multi-wavelength (MWL) campaigns from radio to TeV γ-rays [46], playing a key role in the constraining broadband property from the radio side.

Nevertheless, there is still some outstanding mystery left in the previous VLBI images of this source. First, the core–jet morphology in each lensed image remains extremely stable at least over two decades since the first VLBI images of this system were obtained [48,49]. This implies that the jet speed is either very low or the jet component is associated with a stationary shock that is often claimed in jets of active gamma-ray blazars. Second, while the observed magnification ratio of the lensed images of the core is in agreement with a simple lens model, the observed ratio for the jet component (A2/B2 ≃ 3–3.3) appears to be smaller than the predicted one, implying an additional effect such as "substructure lensing" by a compact foreground clump may be at work.

To better answer these questions, EATING VLBI plays a key role thanks to the higher-resolution and higher-cadense monitoring capability. A preliminary EATING image of B0218+357 shows that both of the lensed images are spatially resolved down to 0.15 mas scales (Figure 6). The jet component in each lensed image was also resolved into substructures. The ongoing multi-epoch analysis of the EATING data will further allow us to constrain the nature of the jet component and its possible motion in great detail.

*4.5. Other Sources*

Besides the sources mentioned above, we are monitoring a growing number of sources with EATING VLBI (e.g., Figure 6). This includes active gamma-ray blazars (e.g., 3C279 and 3C273), nearby radio galaxies (e.g., 3C120 and 3C111), neutrino-emitting blazars (TXS 0506+056), nearby low-luminosity AGN, etc. High-resolution EATING VLBI observations on these sources allow us to resolve the key regions such as the sites of gamma-ray or neutrino production as well as jet collimation/acceleration scales in the vicinity of the central black holes. The EATING VLBI monitoring observations of multiple jet sources at 22 GHz will also play a complementary role in other existing VLBI jet monitoring programs such as the MOJAVE (Monitoring Jets in Active galactic nuclei with VLBA Experiments) at 15 GHz and the Large VLBA Project BEAM-ME at 43/86 GHz (successor to VLBA-BU-BLAZAR) conducted with VLBA by the Boston University Blazar group.

## 5. Future Prospects

As highlighted in the previous sections, EATING VLBI is now in stable operation and the first scientific outcomes are being produced. EATING is a unique VLBI array that allows regular monitoring with global baselines. Nevertheless, the current capability is limited to K-band, 1 Gbps, and single polarization. To take full advantage of EATING, it is desired to update its observing capability.

- **Enhance recording rates:** Increasing the recording rates of the network is essential to enhance the array sensitivity and expand our science targets into weaker objects. Although EATING is formally limited to 1 Gbps, individual regional VLBI networks are already operating at higher recording rates of up to 4–32 Gbps. Commissioning of wideband EATING VLBI observations is currently ongoing.
- **Dual-polarization:** Polarimetric observations are crucial to investigate the magnetic-field properties of relativistic jets. However, some of the primary stations in East Asia have not been capable of dual-polarization until very recently [50]. High-resolution polarimetric EATING will allow us to spatially resolve and monitor the time evolution of the magnetic-field structures in the acceleration and collimation regions of the jets. Dual-polarization is also important to enhance the sensitivity to maser emission.
- **Expand frequency coverage:** While many of the stations in East Asia are capable of observations of up to 43 or 86 GHz bands, compact triple-band (22/43/86 GHz) receivers are being installed at the three Italian telescopes. This will allow us to perform EATING VLBI observations up to 22, 43, and 86 GHz bands quasi-simultaneously, which further increases the angular resolution. An expansion of the EATING frequency coverage to the lower frequency side such as C and L bands is also under consideration.
- **Expansion of network:** Although current EATING has global baselines, they are largely limited to the east–west direction, resulting in highly elongated beam shapes. To expand the north–south baselines, we are currently testing EATING observations in conjunction with telescopes in Australia. We also plan to perform joint EATING observations with Thailand National Radio Telescope in South East Asia, which will nicely fill the UV gap between East Asia and Australia.

## 6. Summary

A collaboration started in 2010 between Italy and Japan in preparation of the VSOP2 launch, which then increased its importance by involving more and more people and institutes in different countries, and by the end of 2012, in a meeting in Bologna, the East Asia To Italy: Nearly Global VLBI (EATING VLBI) collaboration had started.

At present, it is a collaboration among the radio astronomy groups of INAF, NAOJ, KASI, and recently, all institutions associated with EAVN.

Italian and EAVN telescopes are collaborating and observing times can be requested for common observations. A MoU was signed between INAF and KASI for coordinated proposal calls on 9 April 2019. It recognized mutual interest in conducting VLBI experiments, and long-standing collaboration and collaborative research on supermassive black holes. Both parties agreed upon providing 30 h of observation time per semester for common VLBI observations with the INAF telescopes and the KVN. Data are correlated at the Daejeon correlator. Moreover, more observation time can be requested outside the MoU as general scientific activity at the two time allocation committees.

The capability of this 'nearly global' array was shortly shown here but a large increase is expected because of a large technical capability increase: high-frequency receivers in Italian telescopes and a large increase in sensitivity and UV-coverage in East Asia.

The collaboration and friendship inside the EATING VLBI collaboration is expected to increase and it is open to new collaborations.

**Funding:** This research was funded by the Italian Ministry of Foreign Affairs and International Cooperation (MAECI) to increase the scientific collaboration between Italy and Japan; JSPS (Japan Society for the Promotion of Science) Grant-in-Aid for Scientific Research (KAKENHI) (A) 22H00157 (K.H.), (B) 21H01137 (K.H.), 19H01943 (K.H.), JSPS Fund for the Promotion of Joint International Research (Fostering Joint International Research (B)) 18KK0090 (K.H.), and JSPS Grant-in-Aid for Early-Career Scientists 18K13592 (K.H.), the Sumitomo Foundation Grant for Basic Science Research Projects 170201 (K.H.), and the Mitsubishi Foundation 201911019 (K.H.). BWS is grateful for the support from the National Research Foundation of Korea (NRF) funded by the Ministry of Science and ICT (MSIT) of Korea (NRF-2020K1A3A1A78114060). Y.C. is funded by the China Postdoctoral Science Foundation (2022M712084).

**Data Availability Statement:** Images presented in this paper are available under request to the first author.

## Notes

1. http://www.ira.inaf.it/meetings/EatingVLBI/, 15 January 2023
2. http://www.ira.inaf.it/meetings/EatingVLBI/2014/Eating_VLBI_2014/, 15 January 2023
3. https://agn.kasi.re.kr/eatingvlbi/, 15 January 2023
4. https://sites.google.com/a/inaf.it/eating-vlbi-workshop-2019/, 15 January 2023
5. http://www.ira.inaf.it/precise/Home.html, 15 January 2023
6. Typically, at the beginning of April and October, see https://www.radiotelescopes.inaf.it, 15 January 2023
7. https://radio.kasi.re.kr/status_report/status_report.php?site=kvn, 15 January 2023